\documentclass[twocolumn,showpacs,aps,pre]{revtex4}

\newcommand{\BE}{\begin{equation}}
\newcommand{\EE}{\end{equation}}
\newcommand{\BA}{\begin{eqnarray}}
\newcommand{\EA}{\end{eqnarray}}

\begin{document}

\title{Electronic localization in two dimensions}
\author{Zhen Ye}
\affiliation{Wave Phenomena Laboratory, Department of Physics,
National Central University, Chungli, Taiwan 32054}
\date{\today}

\begin{abstract}

By an improved scaling analysis, we suggest that there may appear
two possibilities concerning the electronic localization in two
dimensional random media. The first is that all electronic states
are localized in two dimensions, as already conjectured
previously. The second possibility is that the electronic
behaviors in two and three dimensional random systems are similar,
in agreement with a recent calculation based on a direct
calculation of the conductance with the use of the Kubo formula.
In this case,  non-localized states is possible in two dimensions,
and possess some peculiar properties. A few predictions are
proposed. Moreover, the present analysis seems accommodating
results from previous scaling analysis.

\end{abstract}

\maketitle

Electronic localization refers to situations in which electrons
are confined in space, and the wave function decays exponentially
along any directions\cite{Anderson58,Lee}. According to the
scaling analysis of localization\cite{gang4}, there can be no
metallic state or metal-insulator transition in two dimensions in
zero magnetic field and in the absence of electronic interactions.
In other words, all electrons are always localized in two
dimensions (2D) purely due to multiple scattering by disorders.
This has been the prevailing view for the past twenty years. As
this phenomenon is due to the wave nature of electrons, it is also
called wave localization. Experiments performed in 1980s on
various 2D systems tended to support the prediction of
\cite{gang4} in that there is the expected logarithm increase in
resistivity\cite{Dolan,Bishop,Uren}. The agreement between theory
and experimental observations was encouraging, and for about two
decades, the question of whether a conducting state is possible in
2D was considered resolved, as recently reviewed in Ref.~\cite{EA}

In the last several years, however, experimental evidence started
to appear, indicating that the widely accepted view might not
always have been corrected. Unusual metallic behavior has been
reported in a number of two dimensional electronic
systems\cite{Pudalov,Shashkin,Shashkin2}. An excellent review on
the current developments in the study of localization behavior and
related phenomena in two dimensions was due to Abrahams et
al.\cite{EA}.

As noted by Abrahams et al.\cite{EA}, one of the main observations
to be explained is that the unusual metallic behavior is displayed
down to the near zero temperature under conditions in which 2D
systems are expected to show insulating behavior because of
localization due to disorders, according to the theory of
\cite{gang4}. Various theories have been put forward to explain
this unusual observation, ranging from theory of non-Fermi-liquid
states, superconduting-insulator transition\cite{Goldman}, scaling
theory including electronic interactions\cite{EA2}, percolation
theory\cite{He}, and so on. These theories have been reviewed
critically by Abrahams et al.\cite{EA}. As pointed out by Abrahams
et al., however, while each of the theories is capable of
explaining one or another part of the set of experimental
observations, none of them is able to reconcile all the
experimental results. Although it has been now widely accepted by
the community that the unusual metallic behavior shown in 2D
systems is caused by electronic interactions, which have not been
included in the consideration of the previous
theory\cite{sarachik}, significant disputes remain. For example,
it is pointed out in \cite{Tarasov1} that the previous view on 2D
localization is apparently incomplete and maybe, in the general
case, incorrect. Whether all the electronic states in 2D
disordered media are localized without electronic interactions
therefore still poses an open question\cite{Rama}. This motivates
us to consider further the problem of electronic localization in
2D disordered systems. The question to be addressed here is
whether there could be extended or non-localized states in 2D
random media and these states are {\it purely} caused by
interference of multiple scattering waves, i.~e. not caused by
delocalization effects such as boundary modes, electronic
interactions, random magnetic fields, off-diagonal disorders,
correlated bands, waveguide transmissions\cite{Opt,loc,Corr}.
These effects are known to be able to delocalize or influence the
localization of electrons.

We are naturally led to the question of how to discern the
contradiction between the experimental observation and the
previous assertion that all electronic waves are localized in 2D
disordered systems. In order to resolve this conflict, two
possible approaches may be adopted. One is to find a theory that
can provide a comprehensive picture to all observations. This is
not an easy task and is too ambitious at this stage, mainly
because of complications involved the actual experiments. Often,
various physical effects interplay with each other, making the
data interpretation itself very difficult. The second approach may
be taken by asking why all electrons have to be localized in two
dimensions after all and whether there is some shortcoming in the
previous analysis. These questions are also very difficult and
delicate. But there may still be some hope that the second
approach may be accomplished by looking back at the previous
analysis which has led to the conclusion that all waves are
necessarily localized in 2D. In this approach, it is to verify
what are the conditions that warrant the conclusion. If these
conditions are not satisfied or they could actually obscure the
discernment of the phenomenon of localization, the conclusion from
the analysis may not be applicable. An improvement may thus be
worthwhile.

Here we try to take the second approach to the problem, in the
hope to stimulate further discussions. To proceed, we first repeat
and then discuss the theory in \cite{gang4}. We suggest that the
previous analysis may not be complete and might be unable to
uniquely single out the localization effect, in line with what has
been stated in the introduction of \cite{Tarasov1}. Then we
propose an improved analysis to address the problem of whether all
waves are indeed always localized in 2D systems. We hope to show
that after the improvement not only the results from previous
analysis can be recovered as a possibility, but also there is the
possibility that the localization behavior in 2D actually bears
similarities to that in three dimensions (3D). Specifically, there
could exist the chance that in two dimensions the transition
between localized and non-localized states is possible, by analogy
with 3D. Through the discussion, we perceive that the evident
conflict between the observations and the previous analysis is due
to the difference in the ways that localization is inferred or
interpreted.

While the essence of the current analysis has been very briefly
reported in \cite{CJP}, here I would like to expand the
discussion. For the sake of convenience, some mathematical
derivations are necessarily repeated.

\input epsf.tex
\begin{figure}[hbt]
\begin{center}
\epsfxsize=3.3in \epsffile{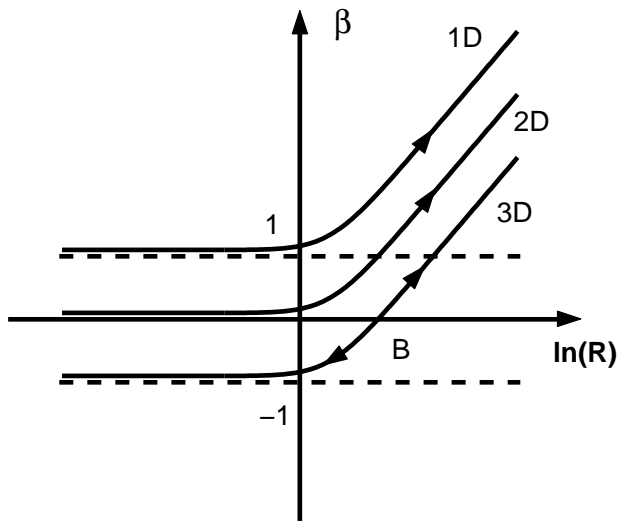} \caption{ \label{fig1}\small
The scaling function $\beta$ versus  $\ln R$ from
Eq.~(\ref{eq:4})}
\end{center}
\end{figure}

First we repeat the previous analysis that leads to the view that
all electrons are localized in 2D, partially for the sake of
convenience on the general reader's part. According to
\cite{gang4}, an hypercubic geometry is used for the scaling
analysis. In the metallic state, the resistance follows the Ohmic
behavior \BE R \sim L^{2-d},\label{eq:1}\EE where $d$ is the
dimension. For a localized state, i.~e. large $R$, the resistance
grows exponentially \BE R\sim e^{L/L_1},\label{eq:2}\EE where
$L_1$ is the localization length which may differ for different
dimensions. A scaling function is defined as \BE \beta =
\frac{\partial \ln R}{\partial \ln L}.\label{eq:3}\EE Taking
Eqs.~(\ref{eq:1}) and (\ref{eq:2}) into (\ref{eq:3}), we obtain
the asymptotic behavior \BE \beta \sim \left\{ \begin{array}{ll}
\ln R, & \mbox{as} \ R\rightarrow \infty \ (\mbox{Localized})
\\ 2-d, & \mbox{as} \ R \rightarrow 0 \ (\mbox{Ohmic})
\end{array}\right. \label{eq:4}
\EE From the asymptotic behavior in Eq.~(\ref{eq:4}), we can
sketch the universal curves in $d=1,2,3$ dimensions. The central
assumptions here are (1) $\beta$ is continuous; (2) $\beta$ is a
function of $R$ and depends on other parameters such as disorders
and length scale only through $R$; and (3) once wave is localized,
the increasing sample size would always mean more localization.
These assumptions have been discussed in some detail in
\cite{gang4}.

The generic behavior of $\beta$ is plotted in Fig.~\ref{fig1}. It
is clear that in the 3D case, the curve crosses the horizontal
axis, yielding an unstable fixed point ($B$). Above this point,
the waves become more and more localized as the sample size
increases. Below the critical point, the system tends to follow
the Ohmic behavior as the sample size is enlarged. This fixed
point separates the localized and non-localized states. For the
two dimensional case, in the Ohmic regime $\beta$ approaches zero
as $\ln(R) \rightarrow 0$. But the perturbation calculation
including the wave interference effect shows that $\beta$ is
always greater than zero. Therefore for both one and two
dimensions, the curves do not cross the horizontal axis, and there
is thus no fixed point. As the sample size increases, all states
move towards the localization regime, as illustrated in
Fig.~\ref{fig1}. This has been the main reason that led previously
to the conclusion that all electronic waves are localized in one
and two dimensions.

The above analysis may require a further consideration. The
reasons follow. Whether a system has non-localized or only
localized states is an intrinsic property of the system, and
should not rely on neither the boundary nor the source\cite{Lee}.
As long as the analysis cannot exclude the possibility that the
boundary or the source is playing a role, the consequence from the
analysis may become questionable. In order to isolate the
localization or non-localization effect, therefore, a genuine
analysis should not be affected by any possible boundary effects
not only in the localization region but also in the
non-localization region. Of course, if the system has indeed only
localized states, the boundary is not an issue, as the dependence
on the boundary is exponentially vanishing\cite{Lee}. However, the
care must be taken for the non-localized regime. It is not
difficult to see that straightly speaking, the above scaling
theory works without ambiguities for situations when both probing
contacts, used to measure the resistance or conductance from which
the localization is inferred, are located outside the sample. In
this case, the Ohmic behavior given by Eq.~(\ref{eq:1}) is valid
under the condition that the current flows uniformly in one
direction. Strictly speaking, this is possible only with (a)
properly scaled sources and with (b) the presence of confining
boundaries. This may be in conflict with the proclamation that
whether it is a localization or non-localized state is the
intrinsic property of the system and should not rely on a boundary
nor a source, and these effects should not come to play a role on
the analysis not only in the localization but also the Ohmic
regions. Thus, in the strictly term, the above analysis may be
more appropriate for studying transport phenomena. It is our
opinion that the reduction in the conductance does not necessarily
mean that all waves are actually localized. In other words, it is
necessary to differentiate the situation that the electrons are
prohibited from transmission from the situation that the system
actually has only localized states. Certainly, we should not
exclude the possibility that the boundary or source effects just
mentioned are minimal and can be practically ignored. In fact,
this possibility is also supported by the following improved
analysis. We also note in reality boundaries are always present in
numerical simulations or experiments. In this case, the influence
of the boundaries has to be carefully addressed. Furthermore, the
earlier scaling analysis has indeed considered the boundaries
\cite{Thouless}.

Taking the view that localization refers to the situation that the
envelope of the wave function decays exponentially from some point
in space\cite{Lee}; any other form of envelope would mean
non-localized states, and also taking the view that whether it is
localized or non-localized is an intrinsic property of the system,
we wish to propose an alternative scaling analysis.

We consider uniformly random systems. Consider an {\it infinite}
system in either 2D or 3D, we compute the effective resistance
between any two space points in the system. Then we investigate
how the resistance varies as the separation ($L$) between the two
points increase. In this way, the boundary or source effect is
unambiguously eliminated. We note that in real experiments, it is
the conductance or the resistance across a sample that is often
measured for transport properties.

In line with the above discussion, in the limit of small
disorders, by neglecting all interference between successive
scatterers the resistance $R$ is assumed to follow the Ohmic
behavior which is guided by $\vec{j} = \sigma \vec{E}$ with
$\sigma$ being the conductivity. An integration leads to \BE R
\sim \left\{\begin{array}{ll} L, & \mbox{for} \ 1D\\\ln(L/L_0), &
\mbox{for}\ 2D\\ \frac{1}{L_0} - \frac{1}{L}, & \mbox{for} \
3D\end{array}\right. \label{eq:5} \EE where $L_0$ refers to the
microscopic size\cite{gang4}. Eq.~(\ref{eq:5}) indicates that the
resistance grows logarithmically either with the sample size or
the distance between two space points in 2D. Hereafter, $L$ can
stand for either the sample size or the separation between two
points in an infinite space.

Eq.~(\ref{eq:5}) is derived by Ohm's law. As the example, we
present the derivations for 2D. Consider an infinite plate, with
two probing contacts at $\vec{r}_1$ and $\vec{r}_2$. One is the
source and the other is the drain. The divergence theorem states
$$ \nabla\cdot \vec{j}(\vec{r}) = -4\pi
A\delta^{(2)}(\vec{r}-\vec{r}_1) + 4\pi
A\delta^{(2)}(\vec{r}-\vec{r}_2). $$ The solution is $$
\vec{j}(\vec{r}) = \frac{A}{|\vec{r}-\vec{r}_1|}\vec{e}_1 -
\frac{A}{|\vec{r}-\vec{r}_2|}\vec{e}_2,  $$ where $$ \vec{e}_1 =
\frac{\vec{r}-\vec{r}_1}{|\vec{r}-\vec{r}_1|}, \ \ \ \vec{e}_2 =
\frac{\vec{r}-\vec{r}_2}{|\vec{r}-\vec{r}_2|}.$$ The coefficient
$A$ is determined from $$ \int |\vec{r}-\vec{r}_1| d\theta
\vec{j}\cdot\vec{e}_1 = I, $$ with $I$ being the current. This
gives \BE A = \frac{I}{2\pi}. \label{eq:A} \EE

Now use Ohm's law $$\vec{j} = \sigma \vec{E}, \ \ \ \mbox{and} \ \
\ \int d\vec{r}\cdot\vec{j} = \sigma \int d\vec{r}\cdot\vec{E}.$$
We integrate from $\vec{r}_1$ to $\vec{r}_2$, but in the
integration we need to exclude the singular point $\vec{r}_1$ and
$\vec{r}_2$. Assume the line between the two points is the
$x$-axis. The integration is $$
A\int_{x_1+\epsilon}^{x_2-\epsilon} dx \frac{1}{x-x_1} +
A\int_{x_1+\epsilon}^{x_2-\epsilon} dx \frac{1}{x_2- x} = \sigma
V. $$ Here the voltage $$ V = \int_{x_1+\epsilon}^{x_2-\epsilon}
dx E(x).$$ The integration gives $$
2A\ln(\frac{L-\epsilon}{\epsilon}) = \sigma V,$$ where $L =
x_2-x_1.$ Considering Eq.~(\ref{eq:A}), we have the resistance $$
\frac{V}{I} = \frac{1}{\pi\sigma} \ln(\frac{L}{\epsilon}). \ \ \
(L
>> \epsilon). $$
Therefore in 2D the resistance scales as $$ R \sim
\frac{1}{\pi\sigma}\ln(\frac{L}{L_0}) \ \ \ (L_0 = \epsilon), $$
which is the result for 2D given in Eq.~(\ref{eq:5}). The result
for 3D can be obtained similarly.

In the other limit where the disorder is very large, the
resistance is large. Therefore an exponential localization is
expected. The resistance is anticipated to grow as \BE R\sim
e^{L/L_1}. \label{eq:6}\EE While the behavior remains unchanged
for 1D, the asymptotic behavior for the scaling function in both
2D and 3D becomes \BE \beta \sim \left\{\begin{array}{ll}
e^{-\ln(R)}, & \mbox{for}\ \ln(R)\rightarrow -\infty;\\ \ln(R), &
\mbox{for} \ \ln(R).\rightarrow
\infty\end{array}\right.\label{eq:7}\EE From these asymptotic
behaviors in the two limits, we expect that the localization
behavior in 2D and 3D should be similar. That is, in both cases,
the scaling function decreases and then increases linearly with
$\ln R$ in the Ohmic and the localization regimes respectively.

\begin{figure}[hbt]
\begin{center}
\epsfxsize=3in \epsffile{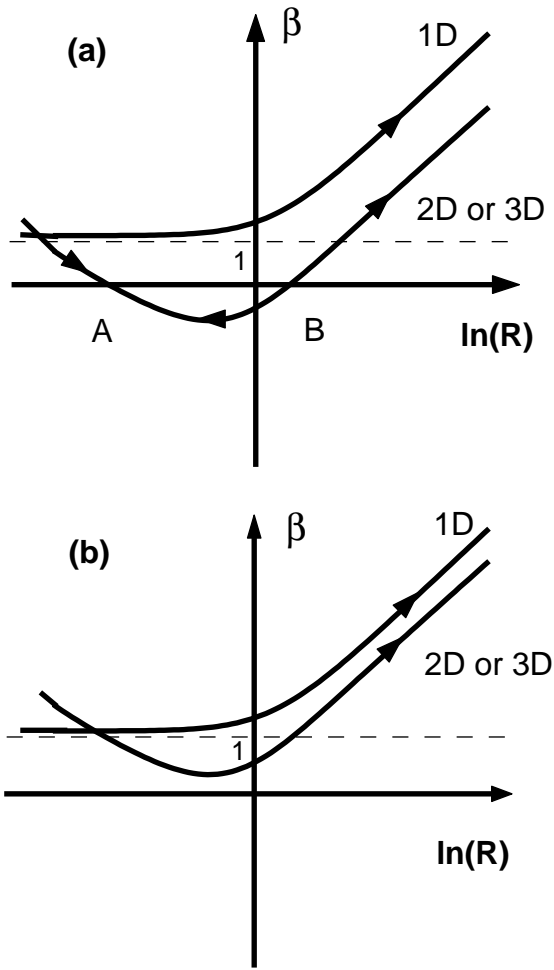} \caption{ \label{fig2}\small
The scaling function $\beta$ vs $\ln(R)$ for 1, 2, and 3
dimensions in the new scaling.}
\end{center}
\end{figure}

Now since we know the asymptotic behaviors in the two limiting
cases (Refer to Eq.~(\ref{eq:7})), the general behavior of the
scaling function $\beta$ may be obtained in the similar way as
outlined above or in \cite{gang4}. Taking the above assumptions
for the scaling function $\beta$ except that it needs not to be
monotonic, the scaling function given by Eq.~(\ref{eq:7}) is
conceptually plotted in Fig.~\ref{fig2}. It is obvious that the 1D
situation is a replicate of that shown in Fig.~\ref{fig1}. The
consequence is that all waves are localized in one dimension for
any given amount of disorders.

Seen from Fig.~2, there are two possibilities for two and three
dimensions. In the first instance shown in Fig.~\ref{fig2}(a),
there are two fixed points: $A$ and $B$. It is clear that $A$ and
$B$ are respectively the stable and unstable fixed points. Point
$B$ separates the localized state and the non-localized state.
When $\ln(R)$ is greater than $B$, the increasing sample size will
lead the system to an infinite resistance in the localized regime;
thus the electronic waves are localized. When $\ln(R)$ is
initially below point $B$, increasing the sample size leads to the
fixed point $A$, where the increasing $L$ will no longer affect
the resistance, indicating a stable non-localized state.

The second possibility is shown in Fig.~\ref{fig2}(b). There is no
fixed point; at most there is only one single unstable fixed
point. In this case, all waves in 2D will be localized like in the
1D situation. This is the case previously considered for 2D.
Previous results affirming that all electronic waves are localized
in 2D fit in this situation. In this sense, the present analysis
accommodates the previous analysis. In this case, the boundary or
source effect mentioned above must be unimportant.

\input epsf.tex

By expanding $\beta$ at point $B$, the critical behavior can be
studied. It is easy to see that at this point, the experimentally
observed symmetry relating conductivity and resistivity could
follow, by analogy with the discussion in \cite{EA2}. Around $B$,
we have \BE \beta \approx a(\ln R - \ln R_B)^r, \ \mbox{with} \ a
> 0,\EE To ensure that point $B$ is unstable, $r$ must be 1. Then
we recover the result in\cite{EA2}\BE R = \left\{
\begin{array}{ll}
 R_B e^{(L/L_B)^a} & \mbox{for}\ R > R_B;\\ R_B
e^{-(L/L_B)^a}& \mbox{for}\ R_B < R.
\end{array}\right.\EE Clearly if $a>1$, the resistance
will grow faster than exponentially. The point $B$ with $R_B$ may
be reached at the limit $L<<L_B$. In this limit, $R/R_B \approx 1
\pm \left(\frac{L}{L_B}\right)^{a}$ for above and below $R_B$
respectively.

To explore the behavior around the stable state at point $A$, we
also expand $\beta$ at this point, \BE \beta \approx -\alpha (\ln
R - \ln R_A)^{\gamma}, \ \mbox{with} \ \alpha
> 0. \EE The fact that point $A$ is a stable point requires $\gamma
= 1$. By integration, we find that \BE R = \left\{
\begin{array}{ll} R_A e^{{(L/L_A)}^{-\alpha}}, & \mbox{for} \ R >
R_A;\\ R_A e^{-{(L/L_A)}^{-\alpha}}, & \mbox{for} \ R_A > R,
\end{array}\right. \label{eq:9}
\EE with $L_A$ is a measure of how fast $R$ approaches $R_A$ and
it may depends on various parameters such as disorders. In the
three dimensional case, when $\alpha = 1$ and $R_A \sim
\frac{1}{\sigma L_0}$, the state at point $A$  is an Ohmic state,
in accordance with Eq.~(\ref{eq:5}). In this case, as $L$
increases, $R$ will converge to the Ohmic result $R_A$. For the
two dimensions, point $A$ refers to a stable state, and when
$\ln(R)$ is initially below $B$ the system will converge to $A$.
The state at point $A$, however, is not of the Ohmic nature. In
fact, if the system were Ohmic, Eq.~(\ref{eq:5}) would imply a
logarithm growth in the resistance with $L$. Therefore, we are led
to the conclusion that although there can be non-localized states
in 2D, the exact Ohmic state is absent, for an infinite sample or
when the separation between two measuring points is exceedingly
large. Due to possible slow convergence of $R$, however, the
resistance may behave nearly like Ohmic for a wide range of $L$.
Furthermore, in 2D, the existence of the stable state at $A$ would
imply that the added small disorder tends to slow down the
increase of resistance near the stable fixed point and the
resistance will be eventually saturated at $A$, as $L$ goes to
infinity, tending to comply with the result shown on P. 203 in
Ref.~\cite{Datta}. This is a state that the classical estimate of
the conductance approaches a minimum conductance due to the wave
nature of electrons, expected from the Landauer formula for the
conductance\cite{Datta,Landauer}. The non-increasing resistance
state may correspond to the unusual metallic behavior discussed in
the literature\cite{EA,EA2}. Whether this state can be observed
depends on samples and experimental conditions. Nonetheless, the
possibly important finding reported here is that a stable
non-localized state is possible in two dimensions, not excluding
the other possibility that two dimensional random media only
support localized states as indicated by Fig.~2(b).

Now we discuss further the above results. First, we pointed out
that the previous scaling analysis has been questioned in the
literature in the past, and more recently by a number of authors
who demonstrated some results in one dimensional random media that
are in contrary to the prediction of the previous
analysis\cite{Lev}. And a novel approach to look into the problem
of localization has been suggested\cite{Genack}. In addition to
the recent experiments\cite{EA}, there were also other experiments
which cannot be accounted for by the previous analysis. As pointed
out in \cite{Asada}, two exceptions predicted by the scaling
theory to this often recited mantra are the extended states which
occur at the center of a Landau level in the quantum Hall
effect\cite{QHE}, and the Anderson transition which occurs in zero
magnetic field if there is a significant spin-orbit
interaction\cite{Asada,Spin1,Spin2,Spin3}.

However, it should also be pointed out that, the previous scaling
analysis was supported by a number of simulations, for instance,
by a simulation of the 2D Anderson model with diagonal
disorders\cite{MacKinnon}. And there were also many experiments
that affirm the previous analysis not only in the electronic
systems (e.~g. \cite{Dolan,Bishop,Uren,Pruisken}) but in phonic
systems\cite{exp}. The fact that the previous analysis was
supported by some research while not by some others may thus imply
that the previous analysis of the 2D localization is incomplete
and is likely model-dependent, as suggested in
\cite{Tarasov1,Lev}. As aforementioned, the present analysis
accommodates the prediction of the previous analysis (Referring to
Fig.~2(b)) and therefore may also accommodate these theoretical
and experimental supports.

Next, we may point out some recent evidence that may support the
present analysis for possible non-localized states (Referring to
Fig.~2(a)). First, our recent exact numerical results indeed
indicate that there is indeed a need to differentiate the
situation that the waves are prohibited from transmission across a
sample from the situation that the system actually has only
localized states\cite{Ye1}. There, the acoustic propagation and
scattering in water containing many parallel air-filled cylinders
is studied in an exact manner. Two situations are compared: (1)
wave propagating through the array of cylinders, imitating common
experimental setups, as summarized in \cite{Genack}, and the
scenario of the previous scaling analysis\cite{gang4}, and (2)
wave transmitted from a source located inside the ensemble. It was
shown that waves can be blocked from propagation by disorders in
the first scenario, but such an inhibition does not necessarily
lead to actual wave localization in the medium. Note that the
electronic system is more complicated because effects such as the
Coulomb interaction can make data interpretation difficult. In
this sense, classical systems are advantageous in studying
localization effects.

Furthermore, the present scaling analysis has been pointed out (Y.
Tarasov, {\it private communication}) to be absolutely in line
with recent findings obtained through direct calculation of the
conductance with the use of the Kubo
formula\cite{Tarasov1,Tarasov2}. In these references, it was shown
that the localization behavior in both 2D and 3D can be similar.

Finally, we note that a suggestion of two fixed points similar to
what was shown in Fig.~2(a) has earlier discussed in a general
description of scaling theories of localization by
Janssen\cite{loc1}. There, however, no concrete physical quantity
was used. We stress that the above inclusion of published research
is far from completeness. There is a great amount of excellent
works which can be referred to in the recent monograph\cite{loc}.

In summary, we have attempted to present an analysis of electronic
localization in random media. The study suggests that the
localization behavior is similar in both two and three dimensions.
The transition between localized and non-localized states is
possible in both dimensions. A new state is predicted as a
possible non-localized state for 2D disordered systems. It must be
stressed that these findings do not exclude the possibility that
all waves are localized in two dimensional random media. As shown
above, the present analysis also supports the possibility that all
waves are localized in two dimensional random media, a profound
principle which has guided significantly previous investigations
of localization since its inception.

This work received support from the National Science Council.


\begin{thebibliography}{99}

\bibitem{Anderson58} P. W. Anderson, Phys. Rev. {\bf 109}, 1492 (1958).

\bibitem{Lee} P. A. Lee and Ramakrishnan, Rev. Mod. Phys. {\bf
57}, 287 (1985).

\bibitem{gang4} E. Abrahams, P. W. Anderson, D. C. Licciardello, and
T. V. Ramakrishnan, Phys. Rev. Lett. {\bf 42}, 673 (1979).

\bibitem{Dolan} G. J. Dolan and D. D. Osheroff, Phys. Rev. Lett.
{\bf 43}, 721 (1979).

\bibitem{Bishop} D. J. Bishop, C. Dynes, and D. C. Tsui, Phys.
Rev. B{\bf 26}, 773 (1982).

\bibitem{Uren} M. J. Uren, R. A. Davies, and M. Pepper, J. Phys.
C{\bf 13}, L985 (1980).

\bibitem{EA} E. Abrahams, S. Kravchenko, and M. P. Sarachik, Rev.
Mod. Phys. {\bf 73}, 251 (2001).

\bibitem{Pudalov} V. M. Pudalov, M. D'Iorio, S. V. Krachenko, and
J. W. Campbell, Phys. Rev. Lett. {\bf 70}, 1866 (1993).

\bibitem{Shashkin} A. A. Shashkin, S. V. Krachenko, and V. T.
Dolgopolov, JETP Lett. {\bf 58}, 220 (1993).

\bibitem{Shashkin2} A. A. Shashkin, S. V. Krachenko, and V. T.
Dolgopolov, Phys. Rev. B{\bf 49}, 14486 (1994).

\bibitem{Goldman} A. M. Goldman and N. Markovic, Phys. Today {\bf
51} (11), 39 (1998).

\bibitem{EA2} V. Dobrosavljevic, E. Abrahams, E. Miranda, and S.
Chakravarty, Phys. Rev. Lett. {\bf 79}, 455 (1997).

\bibitem{He} S. He and X. C. Xie, Phys. Rev. Lett. {\bf 80}, 3324
(1998).

\bibitem{sarachik} M. Sarachik, the lecture at the workshop of Quantum Transport and Mesoscopic
Physics in Hsinchu of Taiwan, January 9-11 (2003).

\bibitem{Tarasov1} Y. Tarasov, J. Phys.: Cond. Matt. {\bf 14}, L357 (2002).

\bibitem{Rama} T. V. Ramakrishnan, Pramana - J. Phys. {\bf 56}, No. 2, 149 (2002).

\bibitem{Opt} Refer to the review articles in {\it Optics of
nanostructured materials}, edited by V. A. Markel and T. F.
George, (Wiley Inter-science, New York, 2001).

\bibitem{loc} M. Janssen, {\it Fluctuations and localization}
(World Scientific, Singapore, 2001); and references therein.

\bibitem{Corr} M. Janssen and K. Pracz, Phys. Rev. E {\bf 61}, 6278
(2000).

\bibitem{CJP} Z. Ye, Chin. J. Phys. {\bf 39}, L207 (2001).

\bibitem{Thouless} D. J. Thouless, {\it Topological quantum
numbers in nonrelativistic physics} (World Scientific, Singapore,
1998).

\bibitem{Datta} S. Datta, {\it Electronic transport in mesoscopic
systems} (Cambridge University Press, New York, 1997).

\bibitem{Landauer} R. Landauer, IBM J. Res. Develop. {\bf 1}, 233
(1957).

\bibitem{Lev} L.I. Deych, D. Zaslavsky, and A.A. Lisyansky, Phys. Rev. Lett. {\bf 81}, 5390 (1998);
L.I. Deych, A.A. Lisyansky, and B.L. Altshuler, Phys. Rev. Lett.
{\bf 84}, 2678 (2000).

\bibitem{Genack} A. A. Chabanov, M. Stoytcher, and A. Z. Genack,
Nature {\bf 404}, 850 (2000).

\bibitem{Asada} Y. Asada, K. Slevin, and T. Ohtsuki, Phys. Rev.
Lett. {\bf 89}, 256601 (2002).

\bibitem{QHE} B. Huckestein, Rev. Mod. Phys. {\bf 67}, 357 (1995).

\bibitem{Spin1} S. Hikami, A. I. Larkin, and Y. Nagaoka, Prog. Theor. Phys.
{\bf 63}, 707 (1980).

\bibitem{Spin2} T. Ando, Phys. Rev. B {\bf 40}, 5325 (1989).

\bibitem{Spin3} Rainer Merkt, Martin Janssen, and Bodo Huckestein, Phys.
Rev. B {\bf 58}, 4394 (1998)

\bibitem{MacKinnon} A. MacKinnon and B. Kramer, Phys. Rev. Lett.
{\bf 47}, 1546 (1981).

\bibitem{Pruisken} H. P. Wei, D. C. Tsui and A. M. M. Pruisken, Phys. Rev. B {\bf 33}, 1488
(1986); A. M. M. Pruisken, in {\it Localization, Interaction and
Transport Phenomena}, edited by B. Kramer, G. Bergmann, and Y.
Bruynseraede, Springer Series in Solid State Sciences, Vol. {\bf
61} (Springer-Verlag, Berlin, 1985), p. 188, and references
therein.

\bibitem{exp} M. van Albada, A. Lagendijk,
Phys. Rev. Lett. {\bf 55}, 2692 (1985); P. E. Wolf, G. Maret,
Phys. Rev. Lett. {\bf 55}, 2696 (1985).

\bibitem{Ye1} Y.-Y. Chen and Z. Ye, Phys. Rev. E {\bf 65}, 056612 (2002).

\bibitem{Tarasov2} Y. Tarasov, J. Low Temp. Phys. ({\it in press}).

\bibitem{loc1} Refer to the discussion for Fig. 5.1 in M. Janssen, {\it Fluctuations and localization}
(World Scientific, Singapore, 2001); and references therein.


\end{thebibliography}
\end{document}